\documentclass[9pt,twocolumn,twoside]{pnas-new}

\templatetype{pnasresearcharticle} 

\usepackage{booktabs} 
\usepackage[T1]{fontenc}
\usepackage{amsmath}
\usepackage{multirow}
\usepackage{array}
\usepackage{color}
\usepackage{listings}
\usepackage{fancyvrb}
\usepackage{cuted}
\usepackage{inputenc}
\usepackage{placeins}
\usepackage{amsmath, amssymb, tabularx}
\usepackage{lipsum}
\usepackage{hyperref}
\usepackage{tabto}
\usepackage{setspace}
\usepackage{balance}
\usepackage{enumitem}
\usepackage{longtable}
\usepackage{pdflscape}
\usepackage{wrapfig}
\usepackage{rotating}
\usepackage{makecell}
\usepackage{array}
\usepackage{pdflscape}
\usepackage{afterpage}
\usepackage{svg}

\usepackage{color}
\usepackage{colortbl}

\definecolor{RowGray}{gray}{0.7}
\definecolor{ColumnGray}{gray}{0.9}

\newcommand\revision[1]{{\color{black}#1}}
\newcommand\minorrevision[1]{{\color{black}#1}}

\title{A Survey on Edge Performance Benchmarking}

\author[a]{Blesson Varghese}
\author[b]{Nan Wang} 
\author[c]{David Bermbach}
\author[d]{Cheol-Ho Hong}
\author[e]{Eyal de Lara}
\author[f]{Weisong Shi}
\author[g]{Christopher Stewart}

\affil[a]{Queen's University Belfast, UK}
\affil[b]{Durham University, UK}
\affil[c]{TU Berlin, Germany, and Einstein Center Digital Future, Germany}
\affil[d]{Chung-Ang University, S. Korea}
\affil[e]{University of Toronto, Canada}
\affil[f]{Wayne State University, USA}
\affil[g]{Ohio State University, USA}

\leadauthor{Varghese} 

\correspondingauthor{\textsuperscript{a}Corresponding e-mail: b.varghese@qub.ac.uk}

\begin{abstract}
Edge computing is the next Internet frontier that will leverage computing resources located near users, sensors, and data stores to provide more responsive services. Therefore, it is envisioned that a large-scale, geographically dispersed, and resource-rich distributed system will emerge and \revision{play a key role in} the future Internet. However, given the loosely coupled nature of such complex systems, their operational conditions are expected to change significantly over time. In this context, the performance characteristics of such systems will need to be captured rapidly, which is referred to as \revision{performance} benchmarking, for application deployment, resource orchestration, and adaptive decision-making. \textit{Edge performance benchmarking} is a nascent research avenue that has started gaining momentum over the past five years. This article first reviews articles published over the past three decades to trace the history of \revision{performance} benchmarking from tightly coupled to loosely coupled systems. It then systematically classifies previous research to identify the system under test, techniques analyzed, and benchmark runtime in edge \revision{performance} benchmarking.
\end{abstract}

\keywords{Edge computing $|$ Edge performance benchmarking $|$ System under test $|$ Quality $|$ Benchmark runtime} 

\dates{This article may be cited as: \textbf{\textit{B. Varghese, N. Wang, D. Bermbach, C. -H. Hong, E. de Lara, W. Shi and C. Stewart, ``A Survey on Edge Performance Benchmarking,'' accepted to ACM Computing Surveys, 2020.}}}

\begin{document}

\maketitle
\thispagestyle{firststyle}
\ifthenelse{\boolean{shortarticle}}{\ifthenelse{\boolean{singlecolumn}}{\abscontentformatted}{\abscontent}}{}


\section{Introduction}
\label{sec:introduction}
The computing landscape has changed significantly over the past three decades. Loosely coupled and geographically dispersed systems have begun replacing tightly coupled monolithic systems~\cite{intro-01}.
One example is from two decades ago, when computing resources that were distributed across numerous organizations and continents were connected under the umbrella of grid computing. Grids offered the unique capability of processing large datasets near data sources without requiring the transfer of data from a distributed workflow to a central system~\cite{intro-06}.
Subsequently, computing became a utility offered remotely through the cloud~\cite{intro-05}.

Although the cloud is the main computing model \revision{adopted} for many Internet-based applications, it has been recognized as an untenable model for the future. This is because billions of devices and sensors are connected to the Internet, and the data generated by these sources cannot be transferred and processed in geographically distant cloud data centers without incurring considerable \revision{communication} delays~\cite{intro-02}. Therefore, the next disruption in the computing landscape is to distribute infrastructure resources and application services further, to bring computing closer to the edge of the network and data sources~\cite{intro-04,paper_pallas_fog4privacy}.
In this article, we use the term ``edge computing'' to refer to the use of resources located at the edge of a network, such as routers and gateways or dedicated micro data centers, to either provide applications with acceleration by co-hosting services in cooperation with the cloud or by hosting them natively \revision{or entirely} on edge resources~\cite{intro-03}.

The inclusion of edge resources for computing creates a large-scale, geographically dispersed, resource-rich distributed system that spans multiple technological domains and ownership boundaries. Such a complex system will be transient, meaning that resources, their availability, and characteristics will change over time. For example, an edge resource previously available for an application may become unavailable based on a recent fault or because the operating system of a target resource may change during maintenance~\cite{intro-08, intro-09}. In this context, it is essential to address the challenge of understanding the relative performance of applications by comparing diverse target hardware platforms from different vendors and their impact on performance when system software changes or new networking protocols are introduced~\cite{intro-07}.
This has motivated the development of \textbf{\textit{edge \revision{performance} benchmarking}}\footnote{\revision{This article will use the terms ``performance benchmarking'' and ``benchmarking'' interchangeably. However, the focus of this article is on edge performance benchmarking.}}~\cite{intro-10,paper_bermbach_fog_vision}.

\revision{Performance} benchmarking is the process of inducing stress on a system while closely observing its responses using a wide range of quality metrics. Typically, synthetic or application-driven workloads are executed on a system under test, such as a virtual machine, a storage system, a stream processing system, or a specific application component, while measuring quality characteristics, such as I/O throughput, end-to-end communication, or computation latency. Unlike alternative approaches such as predictive methods or simulation, insights into real system behaviors can be obtained by replicating the conditions of a production environment~\cite{book_cloud_service_benchmarking}.

\revision{To the best of our knowledge to date, a survey on edge performance benchmarking is not available in the literature.}
Hence, this article focuses on the following aspects: (i) Tracing the history of the development of \revision{performance} benchmarking over the past three decades for high-performance computing (HPC), grid, and cloud systems, (ii) cataloging and examining different edge performance benchmarks, and (iii) reviewing the system under test, techniques analyzed, and benchmark runtime that facilitate edge \revision{performance} benchmarking.

\begin{figure*}[t]
\centering
  \includegraphics[width=0.95\textwidth]{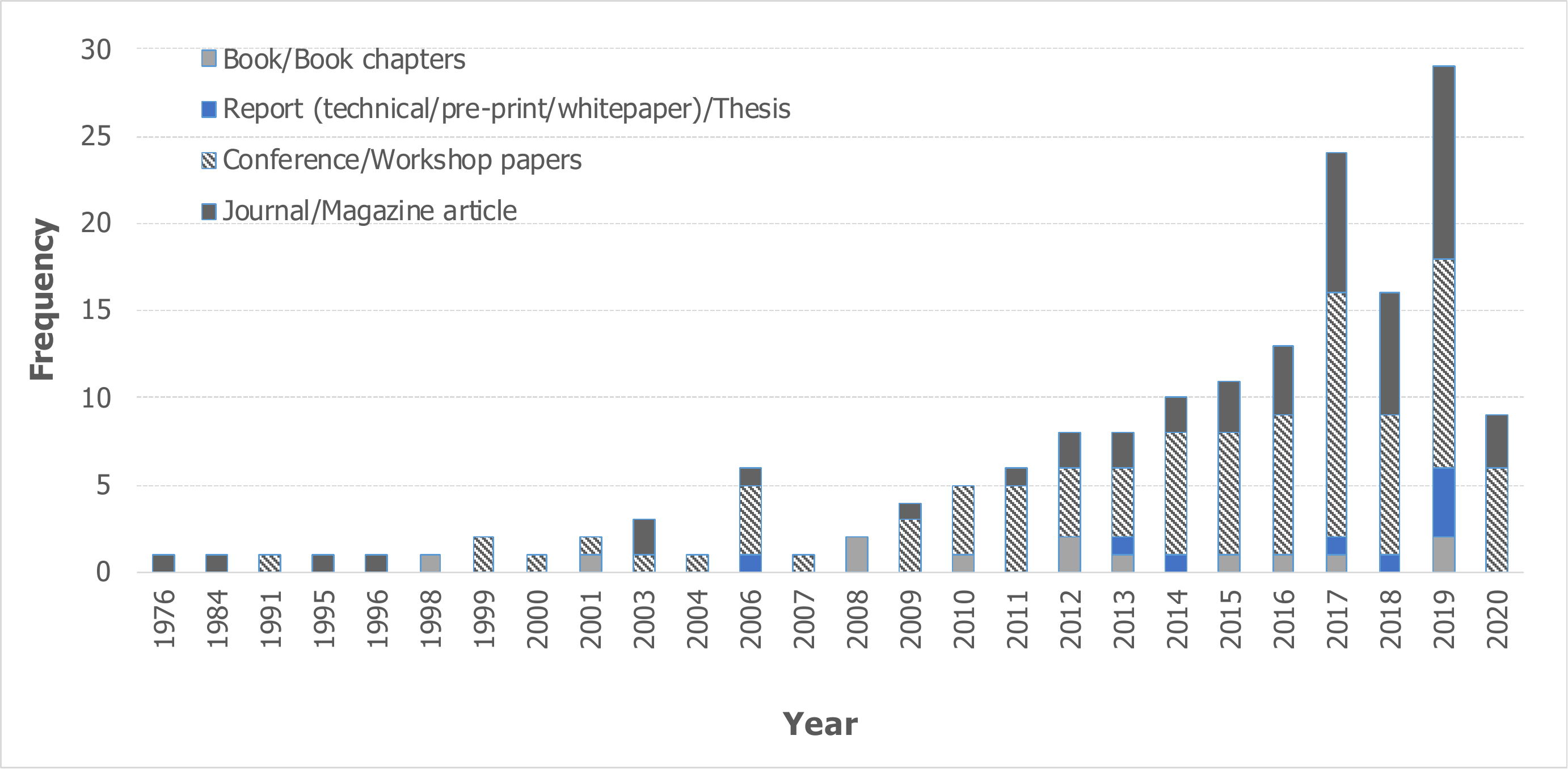}
  \caption{Histogram of the research publications reviewed in this article.}
  \label{fig:histogram-overall}
\end{figure*}

Figure~\ref{fig:histogram-overall} presents a histogram of the total number of research publications reviewed in this article from 1976 and 2020 in the following categories: (i) books and book chapters; (ii) reports, including preprint articles or white papers and doctoral research theses; (iii) conference or workshop papers; and (iv) journal or magazine articles. More than 83\% of the articles reviewed were published after 2010 and more than 61\% of the articles were published after 2015.

\revision{
\subsection{Survey Method}
\label{subsec:surveymethod}

The survey method adopted for preparing this article is based on an approach presented in a previous survey article~\cite{surveymethod-1}. This method includes (1) defining the objectives of the survey, (2) defining research questions, (3) selecting keywords for searching, and (4) identifying criteria for including or excluding research. These aspects are defined below.

(1) The \textit{objectives} of this survey are defined as follows:
\begin{enumerate}
    \item[\textbf{O1}] Provide the research community with a catalog of research related to edge performance benchmarking.
    \item[\textbf{O2}] Trace the development timeline of performance benchmarking for edge systems.
    \item[\textbf{O3}] Understand the key dimensions of existing research on edge performance benchmarking.
    \item[\textbf{O4}] Discuss directions for future research to extend the application of edge performance benchmarking.
\end{enumerate}

(2) The \textit{research questions} addressed by this survey are defined as follows:
\begin{enumerate}
    \item[\textbf{RQ1}] To which systems is edge performance benchmarking applied? This will be discussed in Section~\ref{sec:systemundertest}.
    \item[\textbf{RQ2}] Which techniques are analyzed by edge performance benchmarking? This will be discussed in Section~\ref{sec:techniquesanalysed}.
    \item[\textbf{RQ3}] What are the runtime environments that edge performance benchmarks operate in? This will be discussed in Section~\ref{sec:benchmarkruntime}.
\end{enumerate}

(3) Publication platforms such as the ACM Digital Library, IEEEXplore, ScienceDirect, arXiv, and Google Scholar were considered. The primary \textit{keywords} were a combination of edge (fog, mobile edge, cloud-edge, cloudlet) and benchmarking (benchmark, benchmark suite, micro benchmark, and macro benchmark) with additional keywords such as performance, system, and distributed systems.

(4) The resulting works were screened to identify the most relevant works. 
\minorrevision{A total of 3,764 publications were considered and screened down to 689 publications}.
The initial filters applied were based on the title, followed by the relevance of the abstract. Two types of performance benchmarking works are available, namely explicit and implicit edge performance benchmarking. For research to be selected as explicit performance benchmarking, a benchmarking method, specific benchmark, or toolchain for facilitating performance benchmarking had to be presented, which was determined by inspecting complete papers. Such papers generally contribute to the field of edge benchmarking. \minorrevision{We selected 21 works containing explicit performance benchmarks, which are cataloged in Section~\ref{sec:edgebenchmarking}}.

Implicit performance edge benchmarking works were also included if they presented an evaluation of the performance of a system under testing, technique developed, or runtime, even though such works did not explicitly highlight a benchmark, benchmarking methodology, or benchmarking suite. Studies that did not present a comparative analysis of the system under test, technique developed, or runtime were not considered for implicit benchmarking. The selected works often use novel workloads and measurement approaches that could potentially be incorporated into explicit edge benchmarking research. \minorrevision{A total of 99 implicit performance edge benchmarking publications were selected}. 
Selections based on the above mentioned criteria were validated by at least two of the authors of this article.

This article considers a wide range of papers and Internet sources related to edge performance benchmarking. For this purpose, we focus on explicit and implicit edge performance benchmarks. We do not claim completeness for our selected set of implicit benchmarks for the following two reasons. First, the body of work on edge computing (including closely related topics such as fog computing or mobile edge computing) is very large and cannot be considered within a single survey paper. Second, explicit edge benchmarks can be identified objectively, whereas implicit benchmarks are subjective. Based on careful analysis and validation by at least two authors, the set of selected implicit benchmarks may not necessarily be complete, but it contains no false positives. Therefore, the subset of implicit benchmarks considered within the scope of this article adds value to this survey and to the emerging field of edge performance benchmarking.}

The remainder of this article is organized as follows.
Section~\ref{sec:background} provides a brief history of \revision{performance} benchmarking.
Section~\ref{sec:edgebenchmarking} catalogs different edge \revision{performance} benchmarks.
Section~\ref{sec:systemundertest} presents a review of systems under testing in edge \revision{performance} benchmarking.
Section~\ref{sec:techniquesanalysed} reviews the techniques analyzed in edge \revision{performance} benchmarking.
Section~\ref{sec:benchmarkruntime} surveys runtime execution environments and deployments in edge \revision{performance} benchmarks.
Section~\ref{sec:conclusions} discusses future directions for additional research and concludes this paper.

\section{A Brief Timeline of Performance Benchmarking}
\label{sec:background}
\revision{Performance} benchmarks have played an important role in the evolution of the computing landscape and represent an important field of research and development. 
CPU or processor-related benchmarks have existed since the 1970s. For example, the Whetstone benchmark was developed to measure the floating-point arithmetic performance~\cite{back-01}. Dhrystone is another benchmark developed in the 1980s to evaluate the performance of integers~\cite{back-02}.

\begin{figure*}
    \centering
    \includegraphics[width=\textwidth]{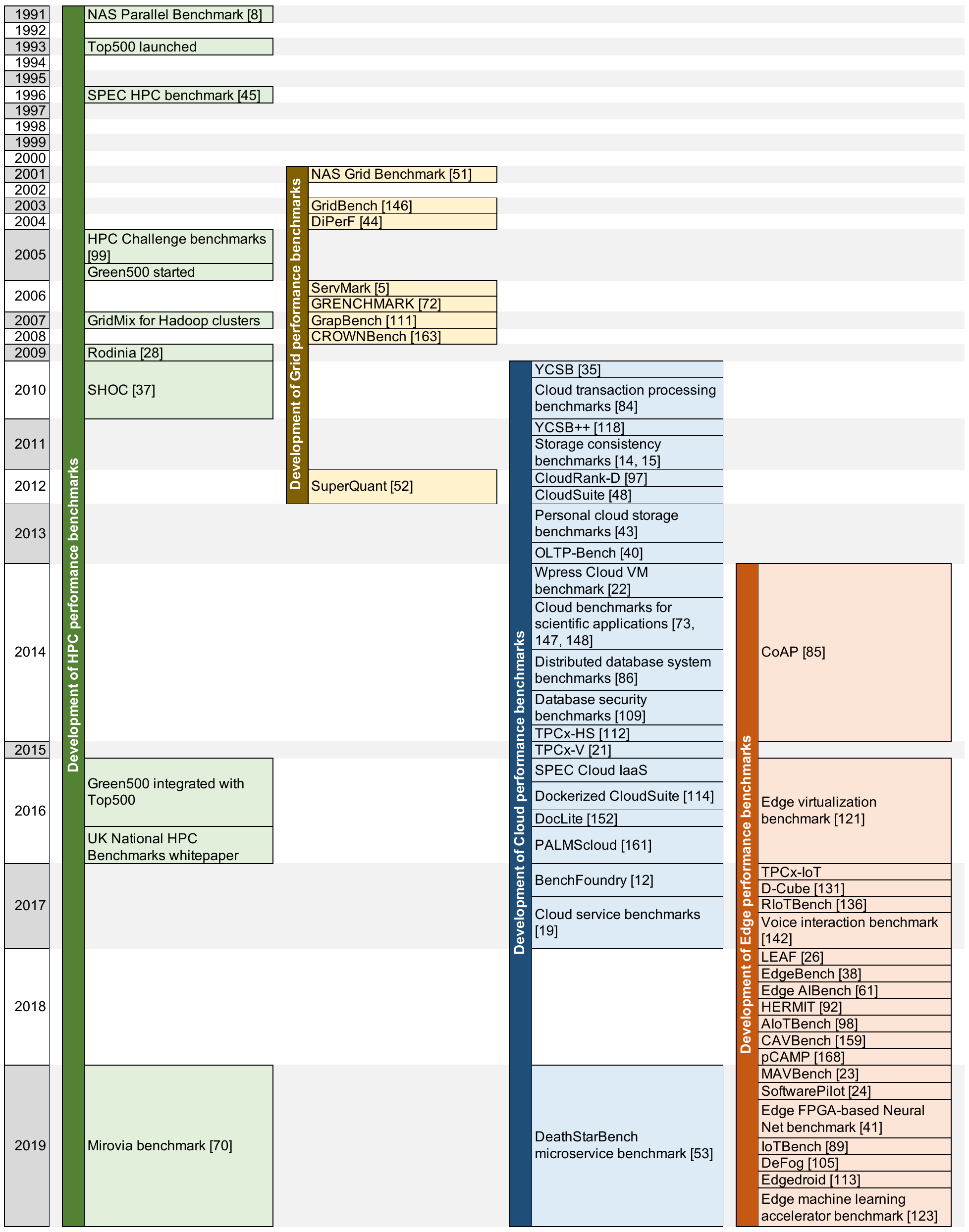}
    \caption{Brief history of the development of performance benchmarking for HPC, grid, cloud and edge systems.}
    \label{fig:backgroundbenchmarks}
\end{figure*}

This section provides a brief history of the development of the field of benchmarking of distributed systems over the past three decades, as shown in Figure~\ref{fig:backgroundbenchmarks}. Specifically, both tightly coupled HPC clusters and supercomputers (highlighted in green), as well as more loosely coupled infrastructure, such as the grid (highlighted in bronze) and the cloud (highlighted in turquoise), are considered. The next section will consider edge computing benchmarks (highlighted in red color), which are the main focus of this article. 

This article does not present an exhaustive timeline of \revision{performance} benchmarking within computer science, but highlights some of the key milestones that have shaped \revision{performance} benchmarking research for current distributed systems and more specifically edge computing, \revision{which is discussed in Section~\ref{sec:edgebenchmarking}}.
In general, the observed pattern is that post-1990 HPC benchmarking, post-2000 grid benchmarking, post-2010 cloud benchmarking, and post-2015 edge benchmarking achieved major milestones. This trend seems to suggest that benchmarks for more loosely coupled distributed systems are gaining prominence as a result of the decentralization and distribution of resources within the computing landscape. 


\subsection{HPC benchmarking}
In 1979, the LINPACK benchmark was developed. This benchmark eventually evolved into the High-Performance LINPACK (HPL) benchmark~\cite{back-03}. This benchmark is computationally intensive and measures the floating point rate of execution by solving a dense system of linear equations. HPL is a de facto benchmark used for capturing the performance of supercomputers and clusters in the TOP500\footnote{\url{https://www.TOP500.org/}} list launched in 1993.
  
The NAS Parallel Benchmarks (NPB) were launched in 1991 and are based on computational fluid dynamics (CFD) applications~\cite{back-04}. 
\revision{The Standard Performance Evaluation Corporation (SPEC) launched several HPC benchmarks in 1996~\cite{back-05}. They focused on three industrial applications, namely seismic processing, 
computational chemistry, 
and climate modeling. 
}

More than a decade after the TOP500 project was launched, energy efficiency became an important metric that influenced the development of parallel computing systems~\cite{back-06}. This resulted in the launch of the Green500\footnote{\url{https://www.TOP500.org/green500/}} project in 2005, which evaluated floating point rates of execution in the context of power consumption. A methodology for measuring and reporting the power used by an HPC system was developed\footnote{\url{https://www.TOP500.org/static/media/uploads/methodology-2.0rc1.pdf}}. In 2016, Green500 was integrated with the TOP500 project. 

The HPC Challenge benchmark was launched in 2005 to measure performance and productivity~\cite{back-07}. This benchmark includes the STREAM benchmark for measuring  sustainable memory bandwidth~\cite{back-09}. 

The field commonly known as big data became popular in the context of HPC clusters, and hence the GridMix\footnote{\url{https://hadoop.apache.org/docs/r1.2.1/gridmix.html}} benchmark for Hadoop clusters emerged in 2007. 
  
The High Performance Conjugate Gradient (HPCG) is a relatively new benchmark that was launched in 2013 to rank supercomputers and clusters in a more balanced manner~\cite{back-10}. The performance of this benchmark is influenced by memory bandwidth. 
 
In 2016, several benchmarks utilized on the UK's national supercomputer ARCHER were presented\footnote{\url{https://www.archer.ac.uk/documentation/white-papers/benchmarks/UK_National_HPC_Benchmarks.pdf}}. These benchmarks are a combination of real applications (DFT, molecular mechanics-based, CFD, and climate modeling) developed by the UK Met Office and synthetic benchmarks from the HPC Challenge. 

Because HPC has become more heterogeneous with the advent of hardware accelerators, such as GPUs, novel  benchmarks have begun to emerge. These benchmarks include the RODINIA~\cite{back-11} and SHOC GPU benchmarks~\cite{back-12}, as well as the more recent Mirovia benchmarks~\cite{back-13}. 

\subsection{Grid Benchmarking}
As academic and research organizations have begun connecting clusters of computers, geographically dispersed and heterogeneous grids have become popular for scientific computing. Key metrics that are relevant to grid benchmarking included turnaround time and throughput because data originated from different geographic locations in a scientific workflow is executed on grids~\cite{back-15}.

One of the first benchmarks for evaluating computing performance on grids was the GridNPB, which was released in 2001. This benchmark is an NPB distribution that uses the Globus grid middleware~\cite{back-14}.

In 2003, GridBench, which allows benchmarks to be specified using a definition language that is compiled into specification languages supported by grid middleware, was released~\cite{back-16}. 

In 2004, the DiPerF benchmarking tool was used to generate performance statistics for grids to predict  grid performance~\cite{back-17}. 

GRENCHMARK was developed for generating synthetic workloads for benchmarking grids~\cite{back-18}. This approach was adopted in ServMark for automating benchmarking pipelines~\cite{back-19}. 

Additional benchmarks such as GrapBench have provided flexibility to benchmarking approaches by considering variations in problems and machine sizes of applications and the grid, respectively~\cite{back-20}. Furthermore, CROWNBench generates synthetic workloads for benchmarking~\cite{back-21}. Domain-specific benchmarks, such as those relevant to financial workloads, have also been developed~\cite{back-22}.

\subsection{Cloud Benchmarking}
Benchmarking or web-based systems, including open, closed, and partially closed systems, have also been considered~\cite{back-35}. \revision{One of the earliest benchmarks for web servers was SPECweb96\footnote{\url{https://www.spec.org/web96/}}}.
In 2010, the first cloud-specific benchmarks were released, namely the Yahoo! Cloud Service Benchmark (YCSB)~\cite{back-23} and transaction processing benchmarks~\cite{back-40}. YCSB++ is an extension of the YCSB that benchmarks scalable column stores~\cite{back-41}.
Various approaches for benchmarking scientific applications in the cloud~\cite{back-24, back-27, back-28} and virtual machines (VMs)~\cite{back-39} have been proposed. Moreover, the aspect of fairness in benchmarking has been considered~\cite{back-29}. Several benchmarks for the consistency of cloud storage services have been developed~\cite{back-36,back-38}.

\revision{A number of cloud performance benchmarks were developed between 2012 and 2015.
Two benchmarks developed by the Transaction Processing Performance Council (TPC) are particularly noteworthy. The first is the TPCx-V~\cite{tpc-01} benchmark, which is a virtual machine benchmark for database workloads. The second is the TPCx-HS~\cite{tpc-02} benchmark, which is a big data benchmark on the cloud.}
Other benchmarks include CloudRank-D~\cite{back-25} and CloudSuite~\cite{back-26} for resources and services. Benchmarks for cloud storage~\cite{back-30,paper_kuhlenkamp_aisle_service_level_pressure}, relational databases~\cite{back-45}, database security~\cite{back-46}, and the scalability and elasticity of distributed databases have also been developed~\cite{back-43}.

\revision{The SPEC benchmarks for infrastructure-as-a-service clouds were launched in 2016 (SPEC Cloud IaaS 2016\footnote{\url{https://www.spec.org/cloud_iaas2016/}})}. At this time, container-based benchmarking suites such as DocLite~\cite{back-31} and a containerized version of CloudSuite~\cite{back-32} also emerged. Moreover, novel cloud hardware architectures could be evaluated using the PALMScloud benchmark suite~\cite{back-33}, and a framework for benchmarking cloud storage services was introduced~\cite{back-37}.
Additionally, a comparison of various benchmarking suites for measuring the quality of cloud services from a client perspective was introduced~\cite{book_cloud_service_benchmarking}. More recently, this has led to the development of benchmarks for microservices on distributed clouds~\cite{back-34,back-44} and benchmarks that can be used in continuous integration processes~\cite{van2012kieker,grambow_continuous_benchmarking:_2019}.

\section{Edge Performance Benchmarking}
\label{sec:edgebenchmarking}
This section highlights developments in edge \revision{performance} benchmarking and then defines \revision{a classification that is used in this article for presenting the different dimensions of research undertaken in the context of edge performance benchmarking.}

As shown in Figure~\ref{fig:backgroundbenchmarks}, edge \revision{performance} benchmarking has been under development since 2015. These benchmarks cover a wide range of resources, including (i) end devices, such as IoT sensors, smartphones, and user gadgets, including wearables; (ii) computational resources located at the edge of a wired network, including routers, switches, gateways, and dedicated resources, including embedded computers and micro clouds; and (iii) cloud resources. Different benchmarks focus on capturing the performance of these resources in both isolated and networked execution contexts.

\begin{figure*}
  \centering
  \includegraphics[width=1.0\textwidth]{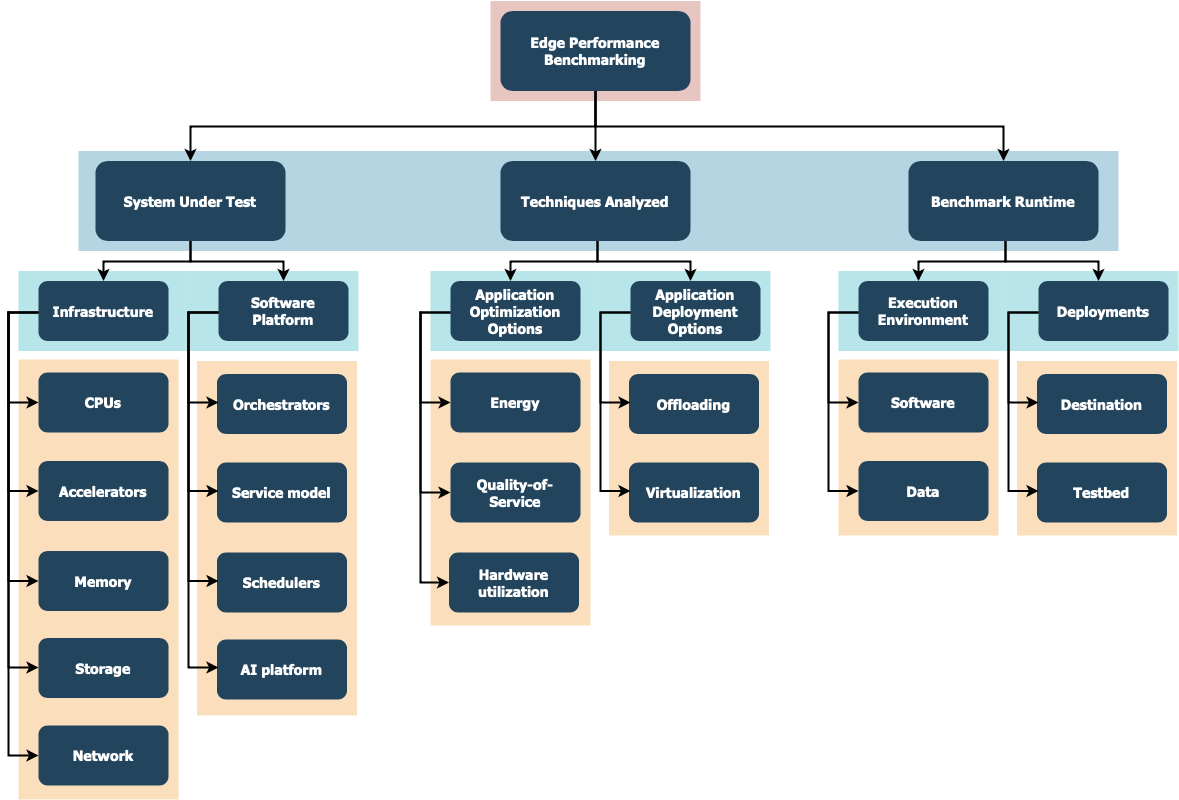}
  \caption{Classification of edge \revision{performance} benchmarking under three dimensions: system under test, techniques analyzed, and benchmark runtime.}
  \label{fig:overall}
\end{figure*}

Existing research on edge performance benchmarking can be classified into the following three dimensions, \revision{as shown in Figure~\ref{fig:overall}}:
\begin{itemize}
    \item \textit{System under test} refers to the hardware infrastructure and software platforms that are benchmarked. This aspect is detailed in Section~\ref{sec:systemundertest} \revision{and aims to address \textbf{RQ1}, which was posed in Section~\ref{subsec:surveymethod}.}
    
    \item \textit{Techniques analyzed} refers to the application optimization options, resource allocation, and application offloading and deployment options analyzed by edge benchmarks, which are discussed in Section~\ref{sec:techniquesanalysed}, \revision{where we aim to address \textbf{RQ2}, which was posed in Section~\ref{subsec:surveymethod}.}
    
    \item \textit{Benchmark runtime} refers to the software and data characteristics of an execution environment and deployment destination, including test beds, which are considered in Section~\ref{sec:benchmarkruntime}, where we \revision{aim to address \textbf{RQ3}, which was posed in Section~\ref{subsec:surveymethod}.}
\end{itemize}

\revision{As highlighted in Section~\ref{subsec:surveymethod}, two types of edge performance benchmarking research can be observed in the literature. The first is explicit performance benchmarking, which is defined as research on developing a benchmarking method, benchmark, or toolchain to facilitate performance benchmarking. The second is implicit performance benchmarking, which refers to research that presents evaluations to capture and compare the performances of any of the aforementioned dimensions (i.e., system under test, techniques analyzed, and benchmark runtime) without specifically presenting a method, benchmark, or toolchain.}

\afterpage{
  \clearpage

\begin{landscape}
\footnotesize
\begin{table}
\caption{List of relevant Edge computing benchmarks, including their type (micro/macro), application domain, benchmarks used, and destination (D - device, E - edge and C - cloud).}
\label{tab:edgebenchmarks}
\begin{tabular}{|c|p{3cm}|p{1cm}|p{3.5cm}|p{10.5cm}|c|}

\hline
\rowcolor{RowGray}
\textbf{Year} & \textbf{Suite/technique} & \textbf{Type} & \textbf{Domain} & \textbf{Benchmarks used} & \textbf{Destination} \\ \hline
\textbf{2014} & CoAP benchmark~\cite{eb-01} & Micro & - & lmbench & D \\ \hline
\textbf{2016} & Virt. benchmark~\cite{eb-02} & Micro & - & NBench, Sysbench, HPL, bonnie++, DD, Stream, Netperf & D \\ \hline
\multirow{6}{*}{\textbf{2017}} & Virt. benchmark~\cite{eb-03} & Micro & - & Sysbench, mbw, fio, iperf, dstat & D \\ \cline{2-6} 
 & Voice benchmark~\cite{eb-04} & Macro & Voice & Mycroft platform & D, E, C \\ \cline{2-6} 
 & \multirow{2}{*}{RIoTBench~\cite{eb-05}} & Micro & \multirow{2}{*}{Stream processing} & 27 IoT steam processing tasks & \multirow{2}{*}{D, C} \\ \cline{3-3} \cline{5-5}
 &  & Macro &  & Transform and load, Statistical summarization, Predictive analytics &  \\ \cline{2-6} 
 & TPCx-IoT\footnote{\url{http://www.tpc.org/tpcx-iot/default5.asp}} & Macro & Power grid & Data ingestion and concurrent queries (based on YCSB) & D, E, C \\ \cline{2-6} 
 & D-Cube~\cite{eb-06} & Micro & Low power systems & 6 wireless protocols & D \\ \hline
\multirow{8}{*}{\textbf{2018}} & CAVBench~\cite{eb-07} & Macro & Autonomous vehicles & SLAM, Object tracking, Battery diagnostics, Speech recognition, Video analytics & E \\ \cline{2-6} 
 & EdgeBench~\cite{eb-08} & Macro & Generic & Speech-to-text, Image recognition, Scalar sensor & E, C \\ \cline{2-6} 
 & Edge AIBench~\cite{eb-09} & Macro & Machine learning & Patient monitor, Surveillance camera, Speech/facial and road sign recognition & D, E, C \\ \cline{2-6} 
 & HERMIT~\cite{eb-10} & Macro & Internet of Medical Things & Physical activity estimation, Advanced encryption standard, Sleep apnea detection, heart rate variability, Histogram equalization, Inverse radon transform, K-means clustering, Lempel-Ziv-Welch compression, Blood pressure monitor & D \\ \cline{2-6} 
 & LEAF~\cite{eb-11} & Macro & Federated learning & Image classification, Sentiment analysis, Text prediction & E \\ \cline{2-6} 
 & \multirow{2}{*}{AIoTBench~\cite{eb-12}} & Micro & 
 Machine learning on mobile devices
 & Neural network layers - convolution, pointwise convolution, depthwise convolution, matrix multiply, pointwise add, ReLU/sigmoid activation, max/avg. pooling & \multirow{2}{*}{D} \\ \cline{3-3} \cline{5-5}
 &  & Macro &  & Image and Speech recognition, Language translation &  \\ \cline{2-6} 
 & pCAMP~\cite{eb-13} & Macro & Machine learning & TensorFlow, Caffe2, MXNet, PyTorch, TensorFlow Lite & D, E, C \\ \hline
\multirow{7}{*}{\textbf{2019}} & DeFog~\cite{intro-07} & Macro & Generic & Object classification, Speech-to-text, Text-audio alignment, Geo-location based mobile game, IoT edge gateway application, Real-time face detection & D, E, C \\ \cline{2-6} 
 & Edgedroid~\cite{eb-15} & Macro & Human-in-the-loop applications & Cognitive assistance application for assembling LEGO & D, E \\ \cline{2-6} 
 & MAVBench~\cite{eb-16} & Macro & Micro aerial vehicles & Scanning, Aerial photography, Package delivery, 3D mapping, Search and rescue & D (Drone), C \\ \cline{2-6} 
 & IoTBench~\cite{eb-17} & Macro & Vision and speech & Video summarization, Stereo image matching, Image recognition, Scan matching, Voice feature extraction, Signals enhancement, Data compression & D \\ \cline{2-6} 
 & SoftwarePilot\footnote{\url{https://www.reroutlab.org/softwarepilot/}}~\cite{boubin2019managing} & Macro & Unmanned aerial vehicle & Aerial photography, Crop surveillance, Rescue and discovery, Facial recognition & E, C\\
 \cline{2-6}
 & Machine learning accelerator benchmark~\cite{eb-18} & Macro & Low power machine learning accelerators & Mobilenet on TensorFlow and OpenVINO & D, E, C \\ \cline{2-6} 
 & Edge FPGA-based Neural Net benchmark~\cite{eb-19} & Macro & Neural network & Keyword spotting application & E \\ \hline
\end{tabular}
\end{table}
\end{landscape}

}


Table~\ref{tab:edgebenchmarks} summarizes \revision{existing explicit performance benchmarking research that is relevant to edge computing systems by considering} the 
benchmark type (micro/macro), application domain, benchmarks used, and destination platforms (device, edge, and cloud) of 21 edge benchmarking techniques or suites. Micro benchmarks refer to benchmarks that capture system-level (CPU, memory, network, storage) performance metrics. By contrast, macro benchmarks refer to benchmarks that capture application-specific system performance metrics for different application domains. Only two benchmarks capture both micro and macro benchmarks, namely RIoTBench~\cite{eb-05} and AIoTBench~\cite{eb-12}. The majority of benchmarks are macro benchmarks and only two utilize generic workloads, namely EdgeBench~\cite{eb-08} and DeFog~\cite{intro-07}. Moreover, only five benchmarks can capture the performance of a complete computation pipeline consisting of a device, edge, and cloud. This can be attributed to the lack of readily available large-scale test beds (although a few are available) for experimentation that integrate a cloud and edge for end users.

The large body of research on edge benchmarking considered in this article relies on either trace data obtained from simulators or on simulators themselves for evaluation, which is contrary to the classic definition of benchmarking. We anticipate experimental benchmarking approaches to be adopted as edge computing matures as a research area and more realistic test beds become readily available. 

\revision{The next three sections will consider both explicit and implicit edge performance benchmarking research and examine this research across the dimensions of system under test, techniques analyzed, and benchmark runtime. Each subsection is organized to discuss explicit edge performance benchmarks first, followed by implicit edge performance benchmarks.}

\section{System Under Test}
\label{sec:systemundertest}
In this section, the systems under testing in edge performance benchmarking are examined by considering two components, infrastructure and software platforms, which are discussed in the following subsections. \revision{Each subsection will discuss both explicit and implicit edge performance benchmarks. The explicit performance benchmarks are listed in Table~\ref{tab:systembenchmarks}.}

\begin{table*}
\revision{
\centering
\footnotesize
\begin{tabular}{|p{7.5cm}|p{0.3cm}|p{0.3cm}|p{0.3cm}|p{0.3cm}|p{0.3cm}|p{0.3cm}|p{0.3cm}|p{0.3cm}|p{0.3cm}|}
\hline
\multirow{3}{*}{} & \multicolumn{5}{c|}{\textbf{Infrastructure}} & \multicolumn{4}{c|}{\textbf{Software}}  \\
 & \multicolumn{5}{c|}{\textbf{}} & \multicolumn{4}{c|}{\textbf{platforms}}  \\
\cline{2-10}
\textbf{System under test characteristics considered by edge performance benchmarks} & \begin{sideways}\textbf{CPUs}~\end{sideways} & \begin{sideways}\textbf{Accelerators}\end{sideways} & \begin{sideways}\textbf{Memory}\end{sideways} &
\begin{sideways}\textbf{Storage}\end{sideways} & \begin{sideways}\textbf{Network}\end{sideways} &
\begin{sideways}\textbf{Orchestrators}\end{sideways} & \begin{sideways}\textbf{Service Model}\end{sideways} & \begin{sideways}\textbf{Schedulers}\end{sideways}  & \begin{sideways}\textbf{AI Platforms}\end{sideways}\\
\hline
CoAP benchmark~\cite{eb-01} &  Y &  N & Y &  N &  N &  N &  N & N & N \\
\hline
Virt. benchmark~\cite{eb-02} &  Y &  N & Y &  Y &  Y &  N &  N & N & N \\
\hline
Virt. benchmark~\cite{eb-03} &  Y &  N & Y &  Y &  N &  N &  N & N & N \\
\hline
Voice benchmark~\cite{eb-04} &  Y &  N & N &  N &  N &  N &  N & N & N \\
\hline
RIoTBench~\cite{eb-05} &  Y &  N & Y &  N &  N &  N &  N & N & N \\
\hline
D-Cube~\cite{eb-06} &  Y &  N & N &  N &  N &  N &  N & N & N \\
\hline
CAVBench~\cite{eb-07} &  Y &  N & Y &  N &  N &  N &  N & N & N \\
\hline
EdgeBench~\cite{eb-08} &  Y &  N & Y &  N &  N &  N &  Y & N & N \\
\hline
Edge AIBench~\cite{eb-09} &  N &  N & N &  N &  N &  N &  N & N & Y \\
\hline
HERMIT~\cite{eb-10} &  Y &  N & Y &  N &  N &  N &  N & N & N \\
\hline
LEAF~\cite{eb-11} &  N &  N & N &  N &  N &  N &  N & N & Y \\
\hline
AIoTBench~\cite{eb-12} &  N &  N & N &  N &  N &  N &  N & N & Y \\
\hline
pCAMP~\cite{eb-13} &  Y &  Y & N &  N &  N &  N &  N & N & N \\
\hline
DeFog~\cite{intro-07} &  Y &  N & N &  N &  Y &  N &  N & N & N \\
\hline
Edgedroid~\cite{eb-15} &  N &  N & N &  N &  N &  Y &  N & N & N \\
\hline
IoTBench~\cite{eb-17} &  Y &  N & N &  N &  N &  N &  N & N & N \\
\hline
Machine learning accelerator benchmark~\cite{eb-18} &  Y &  Y & N &  N &  N &  N &  N & N & N \\
\hline
Edge FPGA-based Neural Net benchmark~\cite{eb-19} &  N & Y & N &  N &  N &  N &  N & N & N \\
\hline
\end{tabular}
\caption{Comparison of the characteristics of system under test in existing edge performance benchmarks}
\label{tab:systembenchmarks}
}
\end{table*}

\subsection{Infrastructure}
\label{subsec:infrastructure}

Embedded CPUs, accelerators, memory, storage hardware, and the network are the infrastructure resources that are considered in edge performance benchmarking.

\subsubsection{CPUs}

\revision{As shown in Table~\ref{tab:systembenchmarks}, the majority of edge performance benchmarks can measure the performance of CPUs used on the edge (five of these benchmarks cannot).}
Many edge benchmarks are evaluated on single-board computers (SBCs), such as Raspberry Pi~\cite{intro-10}, acting as edge resources. An SBC is a circuit board consisting of a CPU, memory, network, storage, and other components. Most SBCs adopt ARM processors as CPUs and have low cost and low power characteristics. Additionally, the performance of modern ARM processors is comparable to that of other general-purpose CPUs~\cite{rajovic2014tibidabo}.

\revision{The CoAP benchmark~\cite{eb-01} utilizes lmbench\footnote{\url{http://lmbench.sourceforge.net/}} for benchmarking ARM processors in Raspberry Pi, BeagleBone, and BeagleBone Black SBCs in terms of bandwidth and latency. The processing overhead of key operations in a modern WSN gateway can also be measured.} An ARM Cortex-A7 dual-core processor hosted by Cubieboard2 was benchmarked~\cite{eb-02} using NBench\footnote{\url{https://nbench.io/}}, sysbench\footnote{\url{https://github.com/akopytov/sysbench}}, and the High-Performance Linpack (HPL) benchmark~\cite{davies2011high}. In a recent study, a wide range of ARM-based SBCs, including the Raspberry Pi 2 model B (RPi2), Raspberry Pi 3 model B (RPi3), Odroid C1+ (OC1+), Odroid C2 (OC2), and Odroid XU4 (OXU4)~\cite{eb-03}, were benchmarked by using sysbench to stress their CPUs. In terms of CPU performance, the Odroid C2 with a Quad-core 2 GHz ARM v8 Cortex-A53 outperformed the other tested SBCs. In terms of power consumption, the Raspberry Pi boards, OC1+, and OC2 achieved high energy efficiency, whereas the OXU4 consumed three to seven times more power than the other SBCs.

\revision{An ARM Cortex A53 CPU hosted by RPi3 was benchmarked~\cite{eb-04} using Mycroft\footnote{\url{https://mycroft.ai/}}, which is an open-source voice assistant. The latency of the pipeline stages of voice interaction was measured, and the results indicated that cloud services outperform RPi, meaning RPi is more suitable for low-complexity tasks compared to the cloud. RIoTBench~\cite{eb-04}, an IoT benchmark for stream processing systems, can measure latency, throughput, jitter, and CPU utilization for VMs that process data flow tasks. D-Cube~\cite{eb-06} is a benchmarking tool that can profile end-to-end delay, reliability, and power consumption for any CPU designed for deploying IoT protocols on the edge.

An Intel Xeon E3-1275 v5 processor included in the Intel fog reference design (FRD) was benchmarked using CAVBench~\cite{eb-07}, which is a benchmarking program for connected and autonomous vehicles. CAVBench deploys computer vision and deep learning applications on a processor and measures the average frames per second and latency. This study concluded that deep learning applications cannot achieve satisfactory performance on the Xeon CPU, meaning accelerators are required. EdgeBench~\cite{eb-08} utilizes audio, image, and scalar pipeline applications to evaluate the computation time, end-to-end latency, and CPU utilization of RPi3. HERMIT~\cite{eb-10} is a benchmarking suite for medical IoT and was utilized to test RPi3 to determine if it is suitable for medical IoT. Various Intel and ARM CPUs were benchmarked using machine learning models in pCAMP~\cite{eb-13}. pCAMP compares TensorFlow, Caffe2, MXNet, PyTorch, and TensorFlow Lite in terms of inference time, total time, and energy consumption. DeFog~\cite{intro-07} was used to evaluate ARM-based SBCs, including RPi3 and OXU4, in various fog computing scenarios consisting of object classification and speech-to-text conversion. IoTBench~\cite{eb-17} offers diverse IoT applications in the vision, speech recognition, and physiological signal processing domains and was used to evaluate RPi3's performance in terms of executed instructions, cycles, and cache misses.}

\subsubsection{Accelerators}
\revision{Although the use of hardware accelerators has been proposed for edge computing~\cite{sut-002}, there is limited use of such accelerators in the explicit edge performance benchmarks listed in Table~\ref{tab:systembenchmarks} (3 of 21~\cite{eb-13, eb-18, eb-19}).}

\revision{pCAMP~\cite{eb-13} benchmarks the inference time of deep learning workloads for edge computing on the Nvidia Jetson TX2.
Low-power and purpose-built accelerators such as the Intel Movidius Myriad XVPU (Vision Processing Unit)~\cite{ionica2015movidius}, Google Edge TPU~\cite{cass2019taking}, NVIDIA 128-core Maxwell and 256-core Pascal architecture-based GPU, and custom field-programmable gate arrays (FPGAs) are benchmarked using deep neural networks on edge devices~\cite{eb-18}.
During the benchmarking process, it was noted that custom FPGAs optimize the performance of specific neural network applications~\cite{eb-19} (also refer to ~\cite{mittal2018survey}).
Various FPGA platforms have been evaluated using custom benchmarks that implement separable convolutional neural network (SCNN) keyword spotting~\cite{benelli2018low}. Their performances were compared to that of the Intel NCS platform in terms of inference time, power consumption, and energy per inference~\cite{eb-19}}.

\revision{In addition to the abovementioned explicit edge performance benchmarks that evaluate accelerators, implicit edge performance benchmarks consider accelerators as well. The most relevant of these benchmarks are presented below.}


The Intel NCS 2 and Google’s Coral USB accelerators were benchmarked using popular inference workloads, namely MobileNet-v1~\cite{howard2017mobilenets} and Inception-v1~\cite{szegedy2015going}, in the MLPerf benchmark~\cite{reddi2019mlperf} in terms of inference time and energy efficiency~\cite{libuttibenchmarking}.

To explore whether a particular deep learning model can provide sufficient accuracy on edge devices, TomoGAN ~\cite{liu2019tomogan, liu2019deep}, which is an algorithm for enhancing the quality of X-ray images, was adapted to run on the Google Edge TPU and NVIDIA Jetson TX2~\cite{abeykoon2019scientific}. The benchmarking results indicated that edge accelerators can provide sufficient accuracy with a novel shallow CNN called the fine-tune network.

In a recent study, the Google Edge TPU, NVIDIA 128-core Maxwell GPU, and Intel Movidius Myriad X VPU were benchmarked by executing eight deep learning models used in personal-scale sensory systems~\cite{antonini2019resource}. The benchmarking results revealed that the Google Edge TPU outperformed the other accelerators on all eight models. In terms of energy efficiency, the Google Edge TPU utilized less than 10 mJ of energy for a single execution of any of the eight models, whereas the other platforms consumed between 5 mJ and 274 mJ depending on the model.

\subsubsection{Memory}

\revision{Seven of the explicit benchmarks in Table~\ref{tab:systembenchmarks} consider memory when benchmarking. The CoAP benchmark~\cite{eb-01} measures memory usage and latency to evaluate the memory performance of SBCs. CAVBench~\cite{eb-07} measures the memory bandwidths and footprints of computer vision and deep learning applications because these applications require significant memory bandwidth and can act as a bottleneck in capacity-limited edge devices. EdgeBench~\cite{eb-08} evaluates the memory utilization of benchmark applications on the Raspberry Pi 3B. HERMIT~\cite{eb-10} was used to analyze the memory characteristics of benchmark applications, and it was determined that a large L1 D-cache and last-level cache (LLC) are required in Raspberry Pi to achieve efficient memory access.}

\revision{A few examples of implicit memory benchmarking in edge computing systems are presented below}. The STREAM benchmark\footnote{\url{https://www.cs.virginia.edu/stream/}}, which measures sustainable memory bandwidth, was used to evaluate edge resources~\cite{eb-02}. The memory performance of copy, scale, add, and triad operations in STREAM was measured on three different software systems: native Linux, Docker, and KVM. The Unix command mbw was utilized to test the memory performance of a wide range of ARM-based SBCs~\cite{eb-03}. The mbw command quantifies available memory bandwidth by transferring large arrays of data in memory. OC1+ outperformed RPi2 because OC1+ adopts 792 MHz LPDDR3 RAM, whereas RPi2 uses 400 MHz LPDDR2 RAM. RPi3 utilizes 900 MHz LPDDR2 RAM and outperforms OC1+ in most cases. Odroid C2 (OC2) and Odroid XU4 (OXU4) outperform RPi2, RPi3, and OC1+ based on their doubled RAM capacity.

\subsubsection{Storage}
\revision{Only two explicit edge performance benchmarks consider storage devices.}
Small form factor edge resources typically use flash-based storage devices, such as embedded MultiMediaCard (eMMC) and MicroSD. The storage performance of edge resources has been evaluated using Bonnie++\footnote{\url{https://www.coker.com.au/bonnie++/}} and the Unix DD command~\cite{eb-02}. Bonnie++ measures data read and write bandwidth, as well as the number of file operations per second. DD is used to measure bandwidth for accessing special device files, such as /dev/zero/.
The fio benchmark\footnote{\url{https://github.com/axboe/fio}} has been used to perform sequential read/write operations in MicroSD cards, and sysbench has been utilized to perform random disk operations on eMMC cards~\cite{eb-03}. The evaluation results revealed that eMMC cards operate at speeds in the order of hundreds of MB/s, whereas MicroSD cards operate in the range of hundreds of Mb/s.

\revision{\subsubsection{Network}
Although the network plays a key role in performance on the edge, there are only two explicit benchmarks that address this aspect.
The network performance of Cubieboard2, which provides a 100BASE-TX Fast Ethernet connection, was evaluated using Netperf\footnote{\url{https://hewlettpackard.github.io/netperf/}} in native, Docker, and KVM environments~\cite{eb-02}. The results revealed that Docker achieves near-native performance, whereas KVM introduces considerable overhead. DeFog~\cite{intro-07} was used to measure communication latency from an edge device to the Amazon cloud when the network bandwidth was fully utilized by the stress-ng operation\footnote{\url{https://kernel.ubuntu.com/git/cking/stress-ng.git/}}.}

\revision{\subsubsection*{Observation \#1}
Even though there have been considerable efforts devoted to benchmarking CPUs and moderate efforts for benchmarking memory, additional effort is still required to measure the edge performance of accelerators, storage, and networks effectively. }

\subsection{Software Platforms}
Orchestrators, cloud services, schedulers, and AI platforms are examples of software platforms that are benchmarked on edge computing systems. Although other software platforms exist, there has been no research in the context of edge performance benchmarking, so these platforms are not discussed in this article.

\subsubsection{Orchestrators}
Orchestrators for edge computing manage edge resources by creating containers, deploying and starting servers, and assigning and scaling computational resources. Orchestration in edge computing is challenging because of limited hardware resources, large volumes of edge resources, and the mobility of connected devices~\cite{intro-10}.

\revision{Edgedroid~\cite{eb-15} is the only explicit edge performance benchmark that considers orchestrators.
Edgedroid evaluates human-in-the-loop applications such as augmented reality and wearable cognitive assistance that are deployed on edge devices. Such applications connect to containers in the cloud for background processing. Edgedroid collects uplink, downlink, and processing time data to identify scaling limits.}

\revision{However, there are also a few implicit edge performance benchmarks that consider orchestrators.}
FocusStack coordinates edge resources for moving targets, such as cars and drones~\cite{ta-59}. This is a challenging task because existing \revision{orchestrators (e.g., OpenStack) were developed} to manage a relatively small number of servers. FocusStack extends OpenStack to support location-based awareness to minimize the number of devices managed at one time. The full-time active monitoring system of FocusStack was benchmarked using the following metric, which represents the total number of bytes transferred every 10 s for monitoring between the orchestrator and a device. FocusStack sent 358 bytes every 10 s while the unmodified OpenStack transferred 17,509 bytes every 10 s.

Edge workload orchestrators that select target edge resources for offloading tasks have been proposed based on fuzzy logic ~\cite{sonmez2019fuzzy}. Such orchestrators can be benchmarked by relying on data obtained from the EdgeCloudSim simulator~\cite{edgecloudsim-01}.
The explored applications include augmented reality, healthcare, intensive computing, and infotainment applications, which are evaluated in terms of service time, failed tasks, and virtual machine utilization.

\subsubsection{Service Model}
\revision{Different service models can facilitate the delivery of services on the edge. Service models can be serverless models or Infrastructure-as-a-Service (IaaS) models. There is limited consideration for such models in the explicit edge benchmarks listed in Table~\ref{tab:systembenchmarks}.
EdgeBench~\cite{eb-08} provides benchmarking applications for the serverless edge computing service model and compares the performances of the AWS IoT Greengrass\footnote{\url{https://aws.amazon.com/greengrass/}} and Azure IoT Edge\footnote{\url{https://azure.microsoft.com/en-us/services/iot-edge/}} platforms.}

\revision{Some implicit edge performance benchmarks consider the IaaS model and two examples are discussed below.}
Nebula implements a decentralized edge cloud by utilizing volunteer edge nodes that provide computational and storage resources~\cite{ryden2014nebula}. Nebula offers the IaaS service model (computation and storage services are available). MapReduce, Wordcount, and InvertedIndex applications have been used to benchmark Nebula in terms of performance, fault tolerance, and scalability~\cite{komosny2015geographic}.

Similarly, FemtoClouds offers IaaS-type services by forming small clusters using smartphones and laptops by leveraging idle or less-loaded resources to fulfill user requests~\cite{habak2015femto}. FemtoClouds has been benchmarked for various metrics, including computational throughput, network utilization, and computational resource utilization.

\subsubsection{Schedulers}
Edge schedulers allow service providers to allocate computing resources efficiently. \revision{None of the explicit edge performance benchmarks can compare the performances of edge schedulers. However, several implicit edge performance benchmarks have considered} resource scheduling policies to satisfy the real-time requirements of smart manufacturing applications in edge computing~\cite{li2019hybrid}. A two-phase scheduling strategy is adopted that first selects a suitable edge computing server based on the target task load, after which additional servers are selected, if necessary, to distribute tasks when one server cannot meet real-time constraints. Schedulers are evaluated on OpenCV applications using metrics such as computing latency, satisfaction degree, and energy consumption.
The fairness aspect of edge scheduling was recently evaluated using synthetic benchmarks~\cite{sut-001}.

\revision{\subsubsection{Artificial Intelligence (AI) Platforms}
Three of the explicit edge performance benchmarks consider AI platforms.
Edge AIBench~\cite{eb-09} provides four AI benchmark applications that can reflect complex scenarios of edge computing, including intensive care unit patient monitoring, surveillance cameras, smart homes, and autonomous vehicles. These test models can be executed using a federated learning framework in a publicly available test bed\footnote{\url{http://www.benchcouncil.org/testbed.html/}}. LEAF~\cite{eb-11} is a modular benchmarking framework for evaluating learning in federated settings. LEAF consists of open-source datasets, statistical and system metrics, and reference implementations. AIoTBench~\cite{eb-12} was designed to evaluate the AI capabilities of edge devices for image classification, speech recognition, transformer translation, and micro workloads.}

\revision{\subsubsection*{Observation \#2}
Explicit edge performance benchmarks do not consider software platforms such as orchestrators, service models, and schedulers, which are vital for performance on the edge. Therefore, existing benchmarks cannot capture the integrated performances of services or applications when different software platforms are adopted.}

\section{Techniques Analyzed}
\label{sec:techniquesanalysed}
This section reviews various techniques that have been analyzed in edge \revision{performance} benchmarks across the dimensions of application optimization and application deployment options. \revision{Table~\ref{tab:techniquesanalysed-existing_benchmark} categorizes the benchmarks listed in Table~\ref{tab:edgebenchmarks}. Table~\ref{tab:techniquesanalysed} summarizes the considered dimensions for a selected set of implicit edge performance benchmarks.}



\begin{table*}
\revision{
\centering
\footnotesize
\begin{tabular}{|p{7.5cm}|p{0.3cm}|p{0.3cm}|p{0.3cm}|p{0.8cm}|p{0.5cm}|}
\hline
\multirow{4}{*}{} & \multicolumn{3}{c|}{\textbf{Application}} & \multicolumn{2}{c|}{\textbf{Application}}  \\
 & \multicolumn{3}{c|}{\textbf{optimization}} & \multicolumn{2}{c|}{\textbf{deployment}}  \\
 & \multicolumn{3}{c|}{\textbf{options}} & \multicolumn{2}{c|}{\textbf{options}} \\
\cline{2-6}
\textbf{Techniques analyzed by edge performance benchmarks} & \begin{sideways}\textbf{Energy consumption}~\end{sideways} &  \begin{sideways}\textbf{Quality-of-Service}\end{sideways} &
\begin{sideways}\textbf{Hardware utilization}\end{sideways} &
\begin{sideways}\textbf{Offloading}\end{sideways}  &
\begin{sideways}\textbf{Virtualization}\end{sideways}\\ 
\hline
CoAP benchmark~\cite{eb-01}, HERMIT~\cite{eb-10} 
 & N & Y & Y  & N & N\\
\hline
Virt. benchmark~\cite{eb-02}
& N & N & Y & N & V, C \\
\hline
Virt. benchmark~\cite{eb-03} &  Y & Y & Y  & N & C \\
\hline
Voice benchmark~\cite{eb-04} &  N &  Y &  N  & D $\xrightarrow[]{}$ E & N \\
\hline
RIoTBench~\cite{eb-05} &  N & Y &  Y  & N & V \\
\hline
TPCx-IoT &  N & Y &  N  & N & N \\
\hline
D-Cube~\cite{eb-06} &  Y &  Y &  N  & N & N \\
\hline
 CAVBench~\cite{eb-07} &  N &  Y &  Y  & N & N\\
\hline
EdgeBench~\cite{eb-08} &  N & Y &  Y  & E $\xrightarrow[]{}$ C & V, C \\
\hline
 Edge AIBench~\cite{eb-09} &  N & N &  N  & N & N \\
\hline
LEAF~\cite{eb-11}, AIoTBench~\cite{eb-12} &  N & N &  N  & N & N \\
\hline
pCAMP~\cite{eb-13} &  Y & Y &  N  & N & N \\
\hline
DeFog~\cite{intro-07} &  N & Y &  N  & C $\xrightarrow[]{}$ E & V, C \\
\hline
Edgedroid~\cite{eb-15} &  N & Y &  N  & N & C \\
\hline
MAVBench~\cite{eb-16} &  Y & Y &  N  & D $\xrightarrow[]{}$ C & N \\ \hline
IoTBench~\cite{eb-17} &  Y & N &  N  & N & N \\ \hline
SoftwarePilot &  N & N &  N  & N & V, C \\ \hline
Machine learning accelerator benchmark~\cite{eb-18} & Y & N &  Y & N & V \\ \hline
Edge FPGA-based Neural Net benchmark~\cite{eb-19} & Y &  Y &  N  & N & N \\ \hline
\end{tabular}
\caption{\revision{Comparison of the characteristics of the techniques analyzed by explicit edge performance benchmarks; Offloading: End-user \textbf{D}evice, \textbf{E}dge, \textbf{C}loud, or \textbf{N}one; Virtualization: \textbf{V}irtual machine, \textbf{C}ontainer, or \textbf{N}one};}
\label{tab:techniquesanalysed-existing_benchmark}
}
\end{table*}

\begin{table*}
\centering
\footnotesize
\begin{tabular}{|p{8.5cm}|p{0.3cm}|p{0.3cm}|p{0.3cm}|p{0.9cm}|p{0.5cm}|}
\hline
\multirow{4}{*}{} & \multicolumn{3}{c|}{\textbf{Application}} & \multicolumn{2}{c|}{\textbf{Application}}  \\
 & \multicolumn{3}{c|}{\textbf{optimization}} & \multicolumn{2}{c|}{\textbf{deployment}}  \\
 & \multicolumn{3}{c|}{\textbf{options}} & \multicolumn{2}{c|}{\textbf{options}} \\
\cline{2-6}
\textbf{Techniques analyzed by edge performance benchmarks} & \begin{sideways}\textbf{Energy consumption}~\end{sideways} & \begin{sideways}\textbf{Quality-of-Service}\end{sideways} 
& \begin{sideways}\textbf{Hardware utilization}\end{sideways}
& 
\begin{sideways}\textbf{Offloading}\end{sideways} & \begin{sideways}\textbf{Virtualization}\end{sideways}\\ 
\hline
Resource allocation benchmark~\cite{ta-02}, MECO benchmark~\cite{ta-01}, RTLBB~\cite{ta-03}, EEDOA~\cite{ta-05} & \multirow{2}{*}{Y}  & \multirow{2}{*}{N} & \multirow{2}{*}{N} &  \multirow{2}{*}{$D\xrightarrow[]{}E$} & \multirow{2}{*}{N}\\
\hline
MIMO benchmark~\cite{ta-68} & \multirow{1}{*}{Y}  & \multirow{1}{*}{N} & \multirow{1}{*}{N} &  \multirow{1}{*}{$D\xrightarrow[]{}C$} & \multirow{1}{*}{N}\\
\hline
IoT benchmark~\cite{ta-26}, LBVS~\cite{ta-22}, MeFoRE~\cite{ta-28}, VEC benchmark~\cite{ta-31} & \multirow{1}{*}{N} &  \multirow{1}{*}{Y} & \multirow{1}{*}{N} & \multirow{1}{*}{$D\xrightarrow[]{}E$} & \multirow{1}{*}{N}  \\
\hline
IoT benchmark~\cite{ta-40} & \multirow{1}{*}{N} & \multirow{1}{*}{Y} & \multirow{1}{*}{N} &  \multirow{1}{*}{$D\xrightarrow[]{}E$} & \multirow{1}{*}{N} \\
\hline
DYVERSE~\cite{intro-09}, ENORM~\cite{intro-08} &  \multirow{1}{*}{N} &  \multirow{1}{*}{Y} &  \multirow{1}{*}{N} &   \multirow{1}{*}{$C\xrightarrow[]{}E$} &  \multirow{1}{*}{V, C} \\
\hline
ECC benchmark~\cite{ta-35} &  N & Y &  N &  $D\xrightarrow[]{}E$ &  N \\
\hline
YEAST~\cite{ta-61}, EDAL~\cite{ta-60} &  Y &  Y &  N  &  N &  N \\
\hline
Content delivery benchmark~\cite{ta-50} &  N &  Y &  Y & N &  N \\
\hline
Cloudlet benchmark~\cite{ta-51} &  N & N &  N  & $C\xrightarrow[]{}E$  &  V \\
\hline
Migration benchmark~\cite{ta-67} &  N & Y &  N  &  $E\xrightarrow[]{}E$ &  C \\
\hline
Cloudlet benchmark~\cite{ta-69} &  Y & Y &  N  & $D\xrightarrow[]{}E$  &  N \\
\hline
Virt. benchmark~\cite{eb-03} &  Y & N &  Y &  N &  C \\
\hline
\end{tabular}
\caption{Comparison of the characteristics of the techniques analyzed by explicit edge performance benchmarks; Offloading: End-user \textbf{D}evice, \textbf{E}dge, \textbf{C}loud, or \textbf{N}one;  Virtualization: \textbf{V}irtual machine, \textbf{C}ontainer, \textbf{N}one}
\label{tab:techniquesanalysed}
\end{table*}

\minorrevision{\subsection{Application Optimization Options}}
\minorrevision{Tables~\ref{tab:techniquesanalysed-existing_benchmark} and \ref{tab:techniquesanalysed} list the different optimization options for target applications that are considered in explicit and implicit edge performance benchmarks, respectively. These options are related to energy consumption, quality of service (QoS), and hardware utilization. 
}

\subsubsection{\revision{Energy Consumption}}
\revision{Energy is an important metric used in many edge performance benchmarks. The virtualization benchmark~\cite{eb-03} compares the power consumption rates of five different SoCs acting as end-user devices. This type of evaluation is useful for estimating the battery durations of end-user devices. To select the best low-power wireless protocol, D-Cube~\cite{eb-06} measures the power consumption of a target edge system while applying different protocols for an IoT application.

Moreover, energy consumption has been employed to evaluate machine learning packages on systems on the device-edge-cloud continuum~\cite{eb-13}. Such benchmarks are useful for optimizing packages for various edge systems. MAVBench~\cite{eb-16} compares the energy consumption of a full-on-edge drone to that of a full-on-cloud drone. This performance benchmark provides a breakdown of the energy consumed by different components of MAVs. Similarly, IoTBench~\cite{eb-17} evaluates the energy dedicated to different components of benchmarks. Recently, several machine learning benchmarks~\cite{eb-18, eb-19} focusing on the energy consumption of accelerators have been developed.

Several implicit edge performance benchmarking studies have compared different energy consumption techniques, a few of which are discussed below.}


Early efforts led to the formulation of a convex optimization problem for minimizing mobile energy consumption~\cite{ta-01}. A multi-user MEC system was considered with a mobile base station and multiple mobile devices, from which tasks were split using a threshold-based policy.

A fine-grained method for multi-resource joint optimization for the energy consumed for task offloading, sub-channel allocation, and CPU-cycle frequency was also developed~\cite{ta-03}. Energy efficiency can also be benchmarked when workloads have varying execution times on MEC servers~\cite{ta-02}.
ThriftyEdge~\cite{ta-04} is used to benchmark the performance of delay-aware task graph partitioning and virtual machine selection to minimize IoT device edge resource occupancy.
Stochastic optimization for minimizing the energy consumption of task offloading while guaranteeing the average queue length of IoT applications was benchmarked in~\cite{ta-05}.

\subsubsection{\revision{Quality-of-Service}}

\revision{QoS is a frequently examined metric in edge performance benchmarks that is represented by execution time, computation and communication latencies, etc. The CoAP benchmark~\cite{eb-01} and HERMIT~\cite{eb-10} measure the execution times of IoT applications on end-user devices.
When comparing application performance across different layers in device-edge-cloud systems, the voice benchmark~\cite{eb-04} and Edgedroid~\cite{eb-15} break down execution times according to application components for analysis. A fine-grained study of application latency under varying edge resource availability characteristics and workloads can be performed using DeFog~\cite{intro-07}. Existing edge performance benchmarks largely focus on hardware with relatively little emphasis on benchmarking different algorithms to optimize edge systems. Therefore, example edge performance benchmarking studies that have focused on optimizing QoS based on algorithm designs are discussed below.}

Because improving application QoS is a key component of edge computing~\cite{intro-02}, the evaluation of QoS is very common in edge literature. We discuss a few examples below. ENORM~\cite{intro-08} benchmarks the benefits of offloading application services from the cloud to the edge based on the QoS of multiple applications on the same edge node.
Priority-based resource scaling approaches can be benchmarked by DYVERSE, which estimates the amount of resources to be added to or removed from a running edge service~\cite{intro-09}. The metric used in this evaluation is the QoS violation rate, which is also employed for evaluating the performance of an optimization model for placing IoT services on edge resources to prevent QoS violations~\cite{ta-40}.


From the perspective of edge service users, MeFoRE~\cite{ta-28} uses previous records of quality of experiences (QoE, such as service give-up ratio) to estimate the resources required by different users. QoE is frequently used in utility functions to optimize the performance of running application services (utility-based optimization). Moreover, the use of QoE to measure user utility for running jobs on the edge versus local execution on mobile devices has been benchmarked~\cite{ta-38}.

\subsubsection{\revision{Hardware Utilization}}
\revision{The utilization of hardware on the edge is another dimension that can be considered to optimize edge performance. This aspect is also captured by edge performance benchmarks. The virtualization benchmark~\cite{eb-02} compares the utilization of different hardware resources on an edge system to select optimal virtualization techniques. Similarly, CPU and memory utilization has been analyzed for benchmarking stream processing platforms~\cite{eb-05} and machine learning accelerators~\cite{eb-18}.

} 

\revision{
\subsubsection*{Observation \#3}
It is noteworthy that the QoS captured by edge performance benchmarks is a dominant criterion for optimizing performance. However, other criteria, such as energy consumption, are not typically captured by edge performance benchmarks.

}

\subsection{Application Deployment Options}
\revision{Edge benchmarks have captured the performance of application deployment options. These include the direction of deployment and determining how to deploy applications on edge systems, which are considered by reviewing edge performance benchmarks for techniques analyzed for computation offloading and virtualization technologies, respectively. The dimension of deployment in the context of the execution environment is considered in Section~\ref{subsec:deployments}.}

\subsubsection{Computation Offloading}

\revision{Only 4 of the 21 explicit edge performance benchmarks listed in Table~\ref{tab:techniquesanalysed-existing_benchmark} consider this dimension. 
This section reviews the computation offloading techniques that are analyzed in explicit edge performance benchmarking.
}

\label{subsubsec:direction}
The following four directions of offloading are relevant to edge environments:

\textit{i. Cloud-to-Edge:} 
\revision{Voice benchmark~\cite{eb-04} characterizes the performance impact of pushing the execution of voice interaction pipelines closer to end-user devices. EdgeBench~\cite{eb-08} offloads serverless functions from the cloud to the edge. Edge AIBench~\cite{eb-09} and DeFog~\cite{intro-07} break down the components of applications and offload some components from the cloud to the edge. Implicit edge performance benchmarks capture the benefits of cloud-to-edge offloading for database replication~\cite{ta-50} and application cloning~\cite{ta-51}.}


\textit{ii. Edge-to-Cloud:} 
This is the typical direction for the internet of connected vehicles, and utility-based multi-level offloading schemes are evaluated to maximize the utility of vehicles, edge servers, and cloud servers~\cite{ta-31}. A collaborative offloading approach has been evaluated to optimize resource allocation and offloading decisions jointly~\cite{ta-65}.

\textit{iii. Edge-to-Edge:}
When an edge node does not have sufficient resources, it can offload (or migrate) its workload to a peer. Two approaches are typically benchmarked: (i) offload forwarding, which forwards all unprocessed workloads to neighboring edge nodes to meet service objectives~\cite{ta-29}; and (ii) service migration, which dynamically migrates services across multiple heterogeneous edge nodes~\cite{ta-35}. Service handoff approaches have been benchmarked to investigate their feasibility for supporting seamless migration to the nearest edge server when a mobile client is moving~\cite{ta-67}.

\textit{iv. Device-to-Edge:} 
\revision{Edgedroid~\cite{eb-15} offloads the backend of task guidance for wearable cognitive assistance from end-user devices to the edge. In addition to existing edge benchmarks}, typical offloading from end-user devices to the edge has been evaluated for applications that require edge data aggregation for energy saving. For aggregating tasks, data from multiple devices are collected by an edge node for preprocessing and filtering tasks~\cite{ta-60, ta-61}.





\subsubsection{\revision{Virtualization Techniques}}
\revision{More than 50\% of the explicit edge performance benchmarks listed in Table~\ref{tab:techniquesanalysed-existing_benchmark} execute applications directly on hardware. Some benchmarks have investigated virtual machines (VMs) and/or containers, and implicit edge performance benchmarking has also considered unikernels.}



\textit{VMs} are important for edge computing in the context of cloudlets~\cite{ta-70}. Implicit benchmarking has focused on the evaluation of VMs on the edge. For example, the selection of the most suitable VM was evaluated for different types of applications~\cite{ta-69}, and a general approach for SLA-driven scheduling for placing VMs in a multi-network operator-sharing edge environment was benchmarked~\cite{ta-22}. \revision{The virtualization benchmark~\cite{eb-02} compares the performances of VMs to those of containers on different edge systems.}



\textit{Containers} have been extensively evaluated for edge systems based on their reduced boot times and lower resource footprints compared to VMs~\cite{ta-71}.
\revision{Among the explicit edge performance benchmarks, the container is the most frequently adopted virtualization technology for application deployment (e.g., EdgeBench~\cite{eb-08}, DeFog~\cite{intro-07}, and Edgedroid~\cite{eb-15}).}
Other benchmarking efforts have highlighted the feasibility of using Docker containers\footnote{\url{https://www.docker.com/}} and Linux containers\footnote{\url{https://linuxcontainers.org/}} as viable options for providing rapid edge deployment~\cite{ta-30, intro-08}.




\textit{Unikernels}\footnote{\url{http://unikernel.org/}} are used for single-purpose applications that use library operating systems and are sealed to modification following deployment~\cite{ta-55}. The corresponding small resource footprints are attractive for edge computing. Unikernels and containers have been benchmarked according to the dimensions of scalability, security, and manageability for IoT applications on the edge~\cite{ta-33}.


\revision{
\subsubsection*{Observation \#4} It is noteworthy that the majority of existing edge performance benchmarks do not capture performance by deploying applications using virtualization techniques. Therefore, such benchmarks can capture the application performance of edge systems that only run a single application, but not those that operate in a multi-tenant edge environment.
}


\section{Benchmark Runtime}
\label{sec:benchmarkruntime}
This section will discuss the execution environments, deployment destinations, and test beds considered by edge \revision{performance} benchmarks. The execution environments highlight the software- and data-related characteristics considered by different edge \revision{performance} benchmarks at run time. Additionally, single and multiple destinations used for deployment by various benchmarks are considered. Finally, different infrastructure deployment options considered by edge performance benchmarks, namely real-world, lab-based, emulated, and simulated test bed infrastructures, are presented.

\subsection{Execution Environment}
\label{subsec:executionenvironment}
Execution environments consist of the software packages and datasets required to execute a benchmark~\cite{burns2001real}. Based on growing support from different programming languages (e.g., Python) and virtualization tools (e.g., Docker), execution environments can be reused across projects, research groups, and scientific disciplines. Open, representative, and comprehensive execution environments broaden research avenues by facilitating scientific exploration in new domains. However, execution environments differ significantly across edge computing applications because such applications change rapidly and frequently. Typically, benchmarks are designed with a narrow focus to target specific workloads by sacrificing the features required to reuse their execution environments.

Table~\ref{tab:benchmarkruntime} lists seven characteristics of open, representative, and comprehensive execution environments. First, software packages and datasets must be accessible to researchers outside the initial study. Clearly, benchmarks are accessible if they use only open-source software (Characteristic \#1). However, benchmarks that represent complex and emergent workloads require custom and/or proprietary software. When it is necessary, such software should be clearly identified and made available (Characteristic \#2). Likewise, the datasets used to drive benchmark execution should be open and easily portable across research contexts (Characteristic \#5).



\begin{table*}
\centering
\footnotesize
\begin{tabular}{|p{0.2cm}|p{4.9cm}|l|l|l|l|l|l|l|l|l|l|l|l|l|l|} 
\hline
\multicolumn{2}{|p{4.9cm}|}{\textbf{Execution environment characteristics of edge benchmarks}}                                               & \begin{sideways}CAVBench~\cite{eb-07}\end{sideways} & \begin{sideways}EdgeDroid~\cite{eb-15}\end{sideways} & \begin{sideways}EdgeBench~\cite{eb-08}\end{sideways} & \begin{sideways}IoTBench~\cite{eb-17}\end{sideways} & \begin{sideways}CoAP benchmark~\cite{eb-01}\end{sideways} & \begin{sideways}MAVBench~\cite{eb-16}\end{sideways} & \begin{sideways}SoftwarePilot~\cite{boubin2019managing}\end{sideways} & \begin{sideways}DeFog~\cite{intro-07}\end{sideways} & \begin{sideways}Edge AIBench~\cite{eb-09}\end{sideways} & \begin{sideways}LEAF~\cite{eb-11}\end{sideways} & \begin{sideways}AIoT Bench~\cite{eb-12}\end{sideways} & \begin{sideways}HERMIT~\cite{eb-10}\end{sideways} & \begin{sideways}pCAMP~\cite{eb-13}\end{sideways} & \begin{sideways}RIoTBench~\cite{eb-05}\end{sideways}  \\ 
\hline
\multirow{4}{*}{\begin{sideways}\textbf{~Software (SW)~}\end{sideways}}   & Reproduced using only open source SW                          & Y                                        & N                                         & N                                        & N                                       & Y                                            & N                                        & N                                       & N                                   & N                                           & Y                                  & N                                       & Y                                    & Y                                   & N                                        \\ 
\cline{2-16}
                                                  & Custom/proprietary
  SW components clearly distinguished and accessible                          & N                                        & N                                         & Y                                        & N                                       & N                                            & Y                                        & Y         & Y                                   & Y                                           & Y                                  & Y                                       & N                                    & N                                   & Y                                        \\ 
\cline{2-16}
                                                  & Employs commercial-grade software                                                             & Y                                        & Y                                         & Y                                        & N                                       & N                                            & Y                                        & Y                                       & Y                                   & Y                                           & N                                  & Y                                       & N                                    & Y                                   & Y                                        \\ 
\cline{2-16}
                                                  & Consider the
  effects of compiler, SW runtime and contextual settings on execution & N                                        & Y                                         & N                                        & N                                       & N                                            & Y                                        & Y                                       & Y                                   & N                                           & Y                                  & Y                                       & N                                    & Y                                   & N                                        \\ 
\hline
\multirow{3}{*}{\begin{sideways}\textbf{Data}\end{sideways}} & Data sets open, accessible
  and portable                                                                     & Y                                        & Y                                         & N                                        & Y                                       & N                                            & Y                                        & Y                                       & Y                                   & Y                                           & Y                                  & Y                                       & Y                                    & Y                                   & Y                                        \\ 
\cline{2-16}
                                                  & Data generation method: \textbf{T}races, \textbf{M}arkov Processes or \textbf{S}imulation~         & T                                        & M                                         & T                                        & T                                       & S                                            & S                                        & M                                       & T                                   & T                                           & M                                  & T                                       & T                                    & T                                   & T                                        \\ 
\cline{2-16}
                                                  & Configure data generation to reproduce a wide range of workload conditions        & N                                        & Y                                         & N                                        & N                                       & N                                            & Y                                        & Y                                       & N                                   & N                                           & Y                                  & N                                       & N                                    & N                                   & N                                        \\
\hline
\end{tabular}
\caption{Comparison of the software and data related execution environment characteristics of edge performance benchmarks}
\label{tab:benchmarkruntime}
\end{table*}

Benchmarks represent real workloads by mimicking specific aspects of their functionality. However, software components with limited functionality place less stress on edge devices compared to commercial-grade software. Therefore, execution environments that are at least partially composed of commercial-grade software can ensure representative demand on edge resources (Characteristic \#3).

Researchers reuse execution environments by adjusting contextual settings to match their target domains. For example, researchers that study edge resources may ignore user actions related to QoS, whereas researchers that study edge-to-cloud offloading may consider multiple contextual settings for QoS. By design, comprehensive benchmarks support a wide range of settings (Characteristic \#7). Execution environments that rely on traces from prior workload executions are often inflexible in terms of runtime and contextual adjustments. Traces have closed data models that cannot be easily extrapolated to what-if conditions. By contrast, execution environments that use data from complex models of simulations or model-driven stochastic methods can be ported to new domains and workload conditions (Characteristics \#4 and \#6).

Table~\ref{tab:benchmarkruntime} reveals several interesting trends. Edge  
benchmarks often employ commercial-grade software components (71\%)
and use open datasets (84\%). These results are likely to be influenced
by the machine learning kernels in edge workloads. Commercial-grade versions
of neural network platforms, object detection models, and speech processors are open sources that are widely used. By contrast, only 50\% of the studied
benchmarks support comprehensive evaluation across runtime and contextual settings.
This result stems from common dependence on traces from prior executions. Furthermore, 75\% of the benchmarks are narrowly tailored to specific contexts and use closed-model traces
that are not easily adapted across runtime and contextual settings.

Among the studied benchmarks, EdgeDroid~\cite{eb-15,wang2019towards},
SoftwarePilot~\cite{boubin2019managing}, and
MAVBench~\cite{eb-16} have the characteristics of open, representative,
and comprehensive execution environments. For example, SoftwarePilot and MAVBench capture virtual reality. Each of these benchmarks employs commercial-grade
open-source software components for machine learning and cognitive assistance.
Additionally, in the corresponding papers, each benchmark is evaluated across
various runtime and contextual settings by adjusting the complexity of machine learning
models or adapting user QoS expectations.
Leaf~\cite{eb-19} is also a comprehensive benchmark that can vary workloads and
runtime settings.
A subtle difference between
these benchmarks is that MavBench relies on simulated environments, which is a design choice
that could yield non-representative workloads if simulations deviate from real-world edge conditions.
By contrast, SoftwarePilot, EdgeDroid, and Leaf use Markov decision models to adapt real
data to new contextual settings. This approach is likely to be more robust for researchers
targeting new domains.

\revision{
\subsubsection*{Observation \#5}
Note that most benchmarks do not operate on data from a real-world edge test bed. Instead, they use simulation data, trace data, or data from a test bed adapted to a different contextual setting. While this issue can be attributed to a lack of readily available edge test beds, it also highlights the need for revisiting and validating existing edge performance benchmarking approaches when new edge test beds become available.

\subsubsection*{Observation \#6}
There is a limited selection of edge performance benchmarks that can generate data to capture a wide range of workload conditions. This factor translates into benchmarks that are narrowly focused on specific applications and cannot be used generically.
}

\subsection{Deployments}
\label{subsec:deployments}
This section will explore the different resource locations at which benchmarks are executed, which are referred to as deployment destinations. 
It will also review test bed options. The following deployment destinations are considered: (i) single destination and (ii) multiple destinations.

\subsubsection{Destination}
As mentioned previously, this article considers the edge as resources located at the edge of a wired network.
However, a number of researchers have also considered user devices as the edge by adhering to a broader definition of the edge (of a network). Therefore, we consider device-only deployment for a single destination.

\textit{i. Single destination} refers to benchmark execution on only one device or edge.

\textit{a. Device-only}: Research that explores device-only benchmarking has been considered for benchmarking in different areas, such as low-power wireless industrial sensors, device-specific protocols, low-overhead virtualization, medical IoT devices, and devices that will execute machine-learning-based workloads.

\textit{Low-power wireless industrial sensors}: Wireless protocols for industrial sensor devices have been benchmarked using observation modules~\cite{eb-06}. The six protocols considered are (i) Enhanced ContikiMAC, (ii) Thompson-sampling-based channel selection, (iii) Glossy, (iv) Chaos, (v) Sparkle, and (vi) Time-slotted channel hopping. Power consumption and end-to-end latencies are profiled, and an additional validation mechanism is incorporated to evaluate the accuracy of benchmarking.

\textit{Device-specific protocols}: Some researchers have benchmarked system architectures for constrained application protocol (CoAP)-based IoT devices~\cite{eb-01}. The key results from benchmarking can be summarized as follows: (i) the selected processor directly impacts CoAP server performance and (ii) the latency of communication channels affects the round-trip times of CoAP requests.


\textit{Low-overhead virtualization}: The performance of virtualization for user devices has been benchmarked extensively~\cite{eb-02, eb-03}. A range of virtualization techniques such as docker containers and kernel-based virtual machines (KVMs) have been considered~\cite{eb-02}. The key results are that hypervisor-based virtualization incurs large overhead and containers seem to be more appropriate for network edges. Container-based virtualization across five different devices has also been considered~\cite{eb-03}. Moreover, there is a negligible impact on performance when using containers compared to bare-metal execution. The characteristics of workloads represent key information for estimating the energy efficiency of devices.

\textit{Medical IoT devices}: The computation and memory characteristics of IoT devices used in the medical domain have been benchmarked~\cite{eb-10}. A collection of medical applications (macro benchmarks) was evaluated against the MiBench, PARSEC, and CPU06 micro benchmarks. The evaluation results revealed that execution characteristics differ between micro and macro benchmarks.

\textit{Devices that run machine learning workloads}: The benchmarking of machine-learning-specific workloads for various devices was presented in~\cite{eb-12}. Both micro benchmarks, such as the individual layers of a neural network, and macro benchmarks, such as applications in image classification, speech recognition, and language translation on the TensorFlow and Caffe2 frameworks, were considered. Similarly, benchmarking for the vision and speech domains has also been considered~\cite{eb-17}.

\textit{b. Edge-only}: Research that explores benchmarking for edge-only deployment will ideally utilize macro benchmarks that are edge native (edge resources are not merely accelerators, but are essential for an application to be used in the real world). Autonomous vehicles are one such application. CAVBench is a benchmarking suite developed for autonomous vehicles by focusing on real-time applications that must process unstructured data~\cite{eb-07}. The edge node employed in this research was designed based on the Intel fog reference design. Hardware accelerators, such as FPGAs on the edge, have been benchmarked for convolutional neural networks in the context of keyword spotting~\cite{eb-19}. It is worth nothing that this use case may not necessarily represent an edge-native benchmark.

\textit{ii. Multiple-destination}: This refers to the execution of a benchmark either across an entire device-edge-cloud resource pipeline or a partial resource pipeline (e.g., device-edge or edge-cloud).

\textit{a. Entire device-edge-cloud pipeline}: There are a number of examples of benchmarking studies that leverage an entire resource pipeline consisting of a device, the edge of a wired network, and a cloud for benchmarking. Three main types of applications are considered.

\textit{Service pushdown}: Voice-interactive applications have been used as macro benchmarks for analyzing entire resource pipelines~\cite{eb-04}. The goal is to push services from the cloud across weak devices and the edge to optimize applications to obtain consistent dialog latency.

\textit{Data aggregation}: Benchmarks that are relevant to the data-intensive workflows of power grids have been presented in the context of an entire resource pipeline (TPCx-IoT)\footnote{\url{http://www.tpc.org/tpc_documents_current_versions/pdf/tpcx-iot_v1.0.4.pdf}}. This workflow enables real-time analytics to be performed on gateways while ingesting data from 200 different types of power station sensors. This benchmark operates in two runs: one run for a warm up and another for measurement. Performance, price, and availability metrics are considered. The majority of benchmarks considered by DeFog fall under this category~\cite{intro-07}.

\textit{Edge inference}: Benchmarks are employed to train machine learning models on the cloud and perform inference on the edge (assuming that a trained model is available on the edge)~\cite{eb-09}. A variety of devices or sensors are considered to generate data in patient monitoring, surveillance, or smart home scenarios. Similarly, inference on a device, edge, or cloud for different machine learning platforms, such as TensorFlow, Caffe2, PyTorch, MXNet, and TensorFlowLite, is considered by pCAMP~\cite{eb-13}, which can also consider different accelerators~\cite{eb-18}. It should be noted that in this case, inference is not distributed because it is performed entirely on the edge.

\textit{b. Partial pipeline}: The following three combinations for benchmarking on partial resource pipelines are considered.

\textit{Device-Edge}: Edgedroid~\cite{eb-15} is an example of benchmarking human-in-the-loop applications, such as cognitive assistance in an edge-cloud deployment, where the edge is a cloudlet. The underlying benchmarking approach is to mimic applications by replaying traces of sensory inputs that are obtained from running the application in the real world. The feedback generated by processing such sensory inputs on the cloudlet is processed using a model of human reactions. This enhances the understanding of latency tradeoffs in the contexts of both the application and edge-cloud deployment.

\textit{Device-Cloud}: The device-cloud pipeline focuses on estimating the performance of distributed intelligence systems~\cite{br-dep-01} and IoT stream application compositions~\cite{eb-05}. The former type of system deploys a sliced neural network across a user device and cloud for distributed inference by estimating where a neural network should be sliced. The latter facilitates benchmarking by combining distributed stream applications using modular IoT tasks.

\textit{Edge-Cloud}: This deployment pipeline is typically used to investigate the performance of applications and the lifecycle of application service offloading from the cloud to the edge and vice-versa (although this is not exclusive to the edge-cloud pipeline and is also relevant in other partial pipelines and the entire resource pipeline (refer to Section~\ref{subsubsec:direction})). Benchmarks that exploit this pipeline include EdgeBench~\cite{eb-08} and SoftwarePilot~\cite{boubin2019managing}.

\revision{
\subsubsection*{Observation \#7}
A number of explicit edge performance benchmarks focus on either a device or edge as a deployment destination. There are relatively few edge performance benchmarks that consider the entire resource pipeline, which integrates a cloud, edge, and device.
}

\subsubsection{Test beds}
The test bed options for deploying edge benchmarks include real-world and physical, lab-based experimental, emulated, or simulated infrastructures.

\textit{i. Real-world physical infrastructure}: This refers to ``in the wild'' physical test beds that closely mimic the operational characteristics of an actual edge deployment. There are only a few such test beds available, such as the Living Edge Lab\footnote{\url{https://www.openedgecomputing.org/living-edge-lab/}} and those reported in~\cite{br-dep-03, br-dep-04}. There has been no evaluation of any of the benchmarks listed in Table~\ref{tab:edgebenchmarks} on real-world infrastructures.

\textit{ii. Lab-based experimental infrastructure}: The majority of test beds on which the edge applications or benchmarks listed in Table~\ref{tab:edgebenchmarks} have been evaluated are lab-based infrastructures. Such test beds may be public cloud offerings or private cloud resources coupled with edge resources in the form of single-board computers~\cite{intro-08, intro-07} (or a cluster~\cite{br-dep-02}) or routers/gateways. End user devices may range from wireless sensors~\cite{eb-06} to user gadgets~\cite{eb-04, intro-08}. Moreover, a number of benchmarks do not rely on real data, but instead use simulation data based on the complexity of the environments they benchmark (e.g., autonomous cars~\cite{eb-07} or drones~\cite{boubin2019managing}).

\textit{iii. Emulated infrastructure}: Real-world physical infrastructure is not easily accessible to many researchers, and many lab-based experimental infrastructures are small in scale and do not represent the characteristics of a real infrastructure. Therefore, various types of emulated infrastructures have recently been proposed~\cite{br-dep-05}.
An emulated environment relies on deploying edge servers on the cloud and configuring these servers, as well as the network and interconnects, such that they represent real edge environments. However, this method assumes knowledge of the parameters required to configure a realistic emulated edge environment, which can only be acquired from a real-world edge infrastructure.

\textit{iv. Simulation infrastructure}: A number of simulators, such as EdgeCloudSim~\cite{edgecloudsim-01}, iFogSim~\cite{ifogsim-01}, FogExplorer~\cite{hasenburg_fogexplorer_2018,hasenburg_supporting_2018}, and MyiFogSim~\cite{myifogsim-01}, are available for edge computing and can provide insights into basic design choices. However, more complex integration tests and application component-specific analysis cannot be performed using such simulators. Therefore, such methods should only be used as a fallback solution when real or emulated benchmarking infrastructures are not available.

\revision{
\subsubsection*{Observation \#8}
Following observation \#5, we note that most edge performance benchmarking is conducted on experimental, emulated, or simulation-based infrastructures that may not be representative of large and geo-distributed real-world physical infrastructures.
}

\section{Future Directions and Conclusions}
\label{sec:conclusions}
\revision{This survey provided a catalog of explicit edge performance benchmarks and a subset of implicit edge performance benchmarking research to achieve objective O1 stated in Section~\ref{subsec:surveymethod}. We presented a brief timeline of performance benchmarking for different computing systems and then considered edge performance benchmarking to achieve objective O2 stated in Section~\ref{subsec:surveymethod}. The key dimensions for edge performance benchmark categorization include the system under test, techniques analyzed, and benchmark runtime, which were examined to achieve objective O3 stated in Section~\ref{subsec:surveymethod}. In exploring the key dimensions of edge performance benchmarks, we addressed the three key research questions posed in Section~\ref{subsec:surveymethod}. Furthermore, we highlighted eight observations relevant to the scope of this paper.
The final objective (O4 in Section~\ref{subsec:surveymethod}) of presenting future research directions is accomplished in this section. The eight observations will be mapped onto future research directions, and the general areas of research will be discussed.
}



\revision{Eight avenues for pursuing future edge performance benchmarking are presented below.}

\textit{i. Widening the scale of geo-distribution}: \revision{Following Observation \#8, most current edge performance benchmarking research is not conducted on real test beds with geo-distributed infrastructures.} Many existing benchmarks are designed to capture the performances of individual cloud or edge resources in lab-based test beds. \revision{Therefore, these benchmarks do not necessarily capture the performance of large-scale geo-distribution that will be observed in an edge computing environment. The absence of comprehensive edge datasets has been also reported previously~\cite{edgedataset-01}.} Although existing performance benchmarks have provided important insights, more comprehensive edge performance benchmarks must fully embrace geo-distributed benchmarking for large-scale collections of cloud and edge resources.

\revision{\textit{ii. Developing edge-specific quality and performance metrics and measurement techniques on different platforms}: As noted in Observation \#2, current edge performance benchmarks do not capture performance on different software platforms, such as orchestrators, schedulers, and service models. To evaluate such software platforms, multiple multi-instance workloads with workflows for the relevant applications are required to test the resource provisioning and sharing capabilities of edge platforms. Addressing this issue would provide researchers with a valuable tool for quantifying the performance of edge applications that will run on a variety of platforms.} Additionally, current efforts largely focus on capturing metrics that are historically relevant to distributed systems, such as grids and clouds. Although such metrics are useful, it is likely that edge-specific metrics considering the transient and massively dispersed nature of such environments have not yet been developed. Additionally, novel techniques to capture these metrics may be required.

\textit{iii. Evaluating benchmarks on real-world infrastructures}: Lab-based infrastructures are the most common test beds used to evaluate edge benchmarks, \revision{as noted in Observation \#5}. At least in part, this can be attributed to limited access to real test beds. Regardless, additional efforts are required to evaluate existing benchmarks on real and resource-rich test beds to identify limitations in current benchmarking approaches.


\revision{\textit{iv. Developing lightweight and rapid edge benchmarks that capture application performance in multi-tenant environments}}: Generally, edge and mobile resources have limited capabilities to execute common and extensive benchmark applications designed for large data center servers. Many HPC applications have been used to capture the performance of CPUs and accelerators for edge computing, but such methods are very time consuming. Additionally, running unmodified Spark or Hadoop applications for big data platforms requires significant time and resources to obtain useful results. Therefore, lightweight benchmarking in terms of actual benchmarks and measurement techniques must be designed and developed for edge platforms. \revision{As noted in Observation \#4, many current edge performance benchmarks execute a single application without considering virtualization. Recent edge systems are designed to utilize virtual machines and containers to support multi-tenancy~\cite{intro-10}. Thus, there is a need to design and develop lightweight and rapid edge benchmarks for multi-tenant edge environments that can quantify the impact of multiple concurrent users competing for the same resources.}

\revision{\textit{v. Developing standardized benchmarks across the entire resource pipeline for capturing offloading performance and varying workload conditions}:} The premise of edge computing is to offload tasks either from a cloud or devices to an edge to reduce the overall response time of an application and improve energy efficiency. \revision{Most edge performance benchmarks do not capture the performance of an entire resource pipeline (devices, edges, and clouds), as noted in Observation \#7, meaning they do not capture the performance of offloading coherently.} Many evaluations presented in the existing literature have used various workloads and metrics relevant to specific platforms or test beds. Therefore, they are non-standard benchmarks and are not compatible across different infrastructures. \revision{Additionally, as noted in Observation \#5, existing edge performance benchmarks have limited flexibility in terms of capturing a wide range of workload conditions.} Hence, a more comprehensive and standard approach for benchmarking offloading mechanisms and varying workload conditions must be considered.

\revision{\textit{vi. Maturing edge performance benchmarking}: Current edge performance benchmarks typically consider QoS metrics, while other relevant criteria, such as energy consumption, are not considered, as noted in Observation \#3. Additionally, most metrics focus on CPUs, but additional research is required to quantify the performance of accelerators, storage, and networks at the edge, as noted in Observation \#1. A benchmarking suite that can holistically capture CPU, accelerator, memory, storage, and network performance on edge platforms and generate performance scores that are normalized against reference platforms is required. These are a few relevant areas that must receive additional attention from the research community to advance edge performance benchmarking research.}

\revision{\textit{vii. Moving beyond performance in edge benchmarking}: In addition to performance in benchmarking, there are other quality dimensions such as elastic scalability, data consistency, security and privacy, and availability that are typically in direct or indirect tradeoff relationships~\cite{diss_bermbach,book_cloud_service_benchmarking}. Current \emph{edge} benchmarking approaches do not consider quality dimensions beyond performance. Benchmarking approaches from other closely related domains, such as cloud computing, can often be adapted or reused for edge environments. For example, there are large numbers of starting points for elastic scalability~\cite{rabl2012solving,back-34,back-40,binnig2009weather,islam2012elasticity,paper_kuhlenkamp_bench_elasticy}, availability~\cite{paper_bermbach_benchmarking_web_apis,paper_bermbach_webapibenchmarking2,shao2009user,toeroe2012service,hauer2020meaningful,fox1999harvest,paper_klems_consistency_benchmarking}, data consistency~\cite{paper_zellag_consistency_benchmarking,paper_wada_consistency_monitoring,paper_golab_analyzing_consistency,paper_anderson_consistency_measurements,back-36,back-38,paper_bermbach_consistency,diss_bermbach}, and security and privacy~\cite{back-46,paper_pallas_et_al_sac_2017_evidence_based,paper_pallas_security_performance_hbase,bermbach2019benchmarking,apostolopoulos1999transport,kant2000architectural,coarfa2006performance,shastri2019understanding}. A more detailed discussion of additional quality dimensions for edge benchmarking is beyond the scope of this paper.}

\textit{viii. Security/privacy-specific edge benchmarks}: Edge systems are significantly more complex than previous iterations of distributed systems (volume of devices connected, heterogeneity of resources and technological domains spanned, and edge resources that are accessible for computing, which were previously unavailable or concealed within networks). This complexity naturally creates a large attack surface and multiple vulnerable spots in terms of data privacy. Therefore, benchmarks that provide insights into identifying security mechanisms for orchestrating services and suitable security standards for complex systems would be very useful.

\revision{\textit{Relevance of edge performance benchmarking to practitioners}: Edge performance benchmarking is a nascent topic within edge computing research. We anticipate that multiple practitioner groups will benefit from edge performance benchmarks. We list four relevant groups below~\cite{intro-07, book_cloud_service_benchmarking}. (i) Edge hardware vendors can tabulate and demonstrate the advantages of edge computing by using performance benchmarks. (ii) System software administrators can investigate the effects on edge applications when changes are introduced within an edge compute infrastructure, such as updates or patches to operating systems, system software, or runtime libraries. (iii) Service providers can select the most appropriate geographic locations for deploying micro and modular data centers on the edge and can quantify the performance of specific applications to justifying their choice of location. (iv) Network administrators may wish to quantify edge application performance when changes are introduced in the network stack, such as a new network protocol or security patch in a specific layer of the stack.

Additionally, real-time edge performance benchmarking can be integrated with automated edge software development and adaptive edge orchestration platforms to select the most appropriate edge resource for deployment based on current performance and network conditions. In this context, edge performance benchmarks will be of significant interest to any edge application developer. }

\acknow{The first author is supported by funds from Rakuten Mobile, Japan and by a Royal Society Short Industry Fellowship.}

\showacknow{} 

\bibliography{references}

\end{document}